\documentclass[11pt]{4ssmmp} 

\usepackage{fancyhdr}    

\usepackage[greek,english]{babel}
\usepackage[iso-8859-7]{inputenc}

\renewcommand{\headrulewidth}{0.4pt}

\begin{document}
 \baselineskip=11pt

\title{Fuzzy Extra Dimensions: Dimensional Reduction,
Dynamical Generation and Renormalizability\hspace{.25mm}\thanks{\,
Based on invited talks presented at ``Noncommutative Spacetime
Geometries" (Alessandria 2007), ``Noncommutative Geometry and
Quantum Spacetime in Physics" (Nishinomiya-Kyoto 2006), ``IV Summer
School in Modern Mathematical Physics" (Belgrade 2006), ``5th Int.
School-Workshop on QFT and Hamiltonian Systems" (Calimanesti 2006),
``Quantum Theory and Symmetries IV" (Varna 2005), ``Supersymmetries
and Quantum Symmetries" (Dubna 2005),  ``BW2005 Workshop" (Vrnjacka
Banja 2005), ``PASCOS 2005" (Gyeongju 2005).}}
\author{\bf{Paolo Aschieri}$^{1,2}$\hspace{.25mm}\thanks{\,e-mail address:
aschieri@theorie.physik.uni-muenchen.de}\ \ \ \bf{Harold
Steinacker}$^3$\hspace{.25mm}\thanks{\,e-mail address:
harold.steinacker@univie.ac.at}\ \ \ \bf{John
Madore}$^4$\hspace{.25mm}\thanks{\,e-mail address:
John.Madore@th.u-psud.fr}\\
\bf{Pantelis Manousselis}$^{5,6}$\hspace{.25mm}\thanks{\,e-mail
address: pman@central.ntua.gr}\ \ \ and \bf{George
Zoupanos}$^5$\hspace{.25mm}\thanks{\,e-mail address:
zoupanos@mail.cern.ch} \vspace{2mm}\\
\normalsize{$^1$ Centro Studi e Ricerche Enrico Fermi,}\\
\normalsize{Compendio Viminale, I-00184, Roma, Italy} \vspace{2mm} \\
\normalsize{$^2$ Dipartimento di Scienze e Tecnologie Avanzate,}\\
\normalsize{Universit{\'a} del Piemonte Orientale, and INFN - Sezione di Torino}\\
\normalsize{Via Bellini 25/G 15100 Alessandria, Italy} \vspace{2mm} \\
\normalsize{$^3$Institut f\"ur Theoretische Physik, Universit\"at
Wien}\\ \normalsize{Boltzmanngasse 5, A-1090 Wien, Austria}
\vspace{2mm} \\
\normalsize{$^4$Laboratoire de Physique
Th\'{e}orique}\\
\normalsize{Universit\'{e} de Paris-Sud,}\\ \normalsize{B\^{a}timent
211, F-91405 Orsay, France}
\vspace{2mm} \\
\normalsize{$^5$ \foreignlanguage{greek}{Τομέας Φυσικής, Εθνικό Μετσόβιο Πολυτεχνείο,}}\\
 \normalsize{15780 \foreignlanguage{greek}{Πολυτεχνειούπολη Ζωγράφου, Αθήνα,} Hellas }
\vspace{2mm} \\
\normalsize{$^6$ \foreignlanguage{greek}{Γενικό Τμήμα, Πολυτεχνική
Σχολή,}}\\ \normalsize{\foreignlanguage{greek}{Πανεπιστήμιο Πατρών,
26110 Πάτρα,} Hellas} }

\date{}

\maketitle

\begin{abstract}
\vspace{.2cm} 

We examine gauge theories defined in higher dimensions where the
extra dimensions form a fuzzy (finite matrix) manifold. First we
reinterpret these gauge theories as four-dimensional theories with
Kaluza-Klein modes and then we perform a generalized \`a la
Forgacs-Manton dimensional reduction. We emphasize some striking
features emerging in the later case such as (i) the appearance of
non-abelian gauge theories in four dimensions starting from an
abelian gauge theory in higher dimensions, (ii) the fact that the
spontaneous symmetry breaking of the theory takes place entirely in
the extra dimensions and (iii) the renormalizability of the theory
both in higher as well as in four dimensions. Then reversing the
above approach we present a renormalizable four dimensional $SU(N)$
gauge theory with a suitable multiplet of scalar fields, which via
spontaneous symmetry breaking dynamically  develops extra dimensions
in the form of a fuzzy sphere $S^2_N$. We explicitly find the tower
of massive Kaluza-Klein modes consistent with an interpretation as
gauge theory on $M^4 \times S^2$, the scalars being interpreted as
gauge fields on $S^2$. Depending on the parameters of the model the
low-energy gauge group can be $SU(n)$, or broken further to $SU(n_1)
\times SU(n_2) \times U(1)$. Therefore the second picture justifies
the first one in a renormalizable framework but in addition has the
potential to reveal new aspects of the theory.
\end{abstract}


\section{Introduction}
In the recent years a huge theoretical effort has been devoted
aiming to establish a unified description of all interactions
including gravity. Out of this sustained endeavor, along with the
superstring theory framework \cite{Green:1987sp}, the
non-commutative geometry one has emerged \cite{Connes, Madore}. An
interesting development worth noting was the observation that a
natural realization of non-commutativity of space appears in the
string theory context of D-branes in the presence of a constant
antisymmetric field \cite{Seiberg:1999vs}, which brought together
the two approaches. However these very interesting approaches do not
address as yet the usual problem of the Standard Model of Elementary
Particle Physics, i.e. the presence of a plethora of free parameters
to the ad-hoc introduction of the Higgs and Yukawa sectors in the
theory. These sectors might have their origin in a
higher-dimensional theory according to various schemes among whose
the first one was the Coset Space Dimensional Reduction (CSDR)
\cite{Forgacs:1979zs, Kapetanakis:hf, Kubyshin:vd}. The CSDR scheme
has been used, among others, to reduce in four dimensions a
ten-dimensional, $N=1, E_8$ gauge theory \cite{Kapetanakis:hf,
Manousselis:2001re, Manousselis:2004xd}and might be an appropriate
reduction scheme of strings over nearly Kaehler manifolds
\cite{Manousselis:2005xa}. The dimensional reduction of gauge
theories defined in higher dimensions where the extra dimensions
form a fuzzy coset (a finite matrix manifold) has been examined in
\cite{Aschieri:2003vy, Aschieri:2004vh}.

In the  CSDR one assumes that the form of space-time is $M^{D}=M^{4}
\times S/R$ with $S/R$ a homogeneous space (obtained as the quotient
of the Lie group $S$ via the Lie subgroup $R$). Then a gauge theory
with gauge group $G$ defined on $M^{D}$ can be dimensionally reduced
to $M^{4}$ in an elegant way using the symmetries of $S/R$, in
particular the resulting four-dimensional gauge group is a subgroup
of $G$. Although the reduced theory in four dimensions is power
counting renormalizable the full higher-dimensional theory is
non-renormalizable with dimensionful coupling. The CSDR scheme
reduces dimensionally a gauge theory with gauge group $G$ defined on
$M_4 \times S/R$ to a gauge theory on $M_4$ imposing the principle
that fields should be invariant under the $S$ action up to a $G$
gauge transformation. The CSDR scheme constitutes an elegant and
consistent truncation of the full theory in four dimensions, keeping
only the first terms of the field expansion in higher harmonics of
the  compact coset spaces. When keeping all the higher harmonics,
i.e. the Kaluza-Klein modes in four dimensions, then the theory in
general becomes non-renormalizable as expected, since the theory was
originally defined in higher than four dimensions. Still it is very
interesting the fact that one can discuss the dependence of the
couplings of the theory on the cutoff, or the beta-function of the
couplings in the Wilson renormalization scheme \cite{Dienes:1998vg,
Kubo:1999ua, Kubo:2000hy}. In the fuzzy-CSDR we apply the CSDR
principle in the case that the extra dimensions are a finite
approximation of the homogeneous spaces $S/R$,  i.e. a fuzzy coset.
Fuzzy spaces are obtained by deforming the algebra of functions on
their commutative parent spaces. The algebra of functions (from the
fuzzy space to complex numbers) becomes finite dimensional and
non-commutative, indeed it becomes a matrix algebra. Therefore,
instead of considering the algebra of functions $Fun(M^{D})\sim
Fun(M^{4}) \otimes Fun(S/R)$ we consider the algebra $A= Fun(M^{4})
\otimes Fun((S/R)_F)$ where $Fun(M^{4})$ is the usual commutative
algebra of functions on Minkowski space $M^{4}$ and
$Fun((S/R)_F)=M_N$ is the finite dimensional non-commutative algebra
of $N\times N$ matrices that approximates the functions on the coset
$S/F$. On this finite dimensional algebra we still have the action
of the symmetry group $S$; this very property allows us to apply the
CSDR scheme to fuzzy cosets. The reduction of a gauge theory defined
on $M^4 \times (S/R)_F$ to a gauge theory on $M^4$ is a two step
process. One first rewrites the higher-dimensional fields, that
initially depends on the commutative coordinates $x$ and the
noncommutative ones $X$, in terms of only the commutative
coordinates $x$,  with the fields now being also $N\times N$ matrix
valued. One then imposes the fuzzy-CSDR constraints on this
four-dimensional theory. We can say that the theory is a
higher-dimensional theory because the fuzzy space $(S/R)_F$ is a
noncommutative approximation of the coset space $S/R$; in particular
the spatial symmetry group $S$ of the space $(S/R)_F$ is the same as
that of the commutative space $S/R$. However the noncommutative
theory has the advantage of being power counting renormalizable
\cite{Aschieri:2005wm} because $Fun((S/R)_F)$ is a finite
dimensional space; it follows that also after applying the
fuzzy-CSDR scheme we obtain a power counting renormalizable theory.
The specific example of the fuzzy sphere is presented.

Next we reverse the above approach \cite{Aschieri:2006uw} and
examine how a four dimensional gauge theory dynamically develops
higher dimensions. The very concept of dimension
therefore gets an extra, richer dynamical perspective. We present a
simple field-theoretical model which realizes the above ideas. It is
defined as a renormalizable $SU(N)$ gauge theory on four dimensional
Minkowski space $M^4$, containing 3 scalars in the adjoint of
$SU(N)$ that transform as vectors under an additional global $SO(3)$
symmetry with the most general renormalizable potential.  We then
show that the model dynamically develops fuzzy extra dimensions,
more precisely a fuzzy sphere $S^2_{N}$. The appropriate
interpretation is therefore as gauge theory on $M^4 \times S^2_{N}$.
The low-energy effective action is that of a four dimensional gauge
theory on $M^4$, whose gauge group and field content is dynamically
determined by compactification and dimensional reduction on the
internal sphere  $S^2_{N}$.  An interesting and  rich pattern of
spontaneous symmetry breaking appears, breaking the original $SU(N)$
gauge symmetry down to
 either $SU(n)$ or $SU(n_1)\times
SU(n_2) \times U(1)$. The latter case is the generic one, and
implies also a monopole flux induced on the fuzzy sphere. The values
of $n_1$ and $n_2$ are determined dynamically.

We find moreover explicitly the tower of massive Kaluza-Klein modes
corresponding to the effective geometry, which justifies the
interpretation as a compactified higher-dimensional gauge theory.
Nevertheless, the model is renormalizable.

A similar but different mechanism of dynamically generating extra
dimensions has been proposed some years ago in
\cite{Arkani-Hamed:2001ca}, known under the name of
``deconstruction''.  In this context, renormalizable four
dimensional asymptotically free gauge theories were considered,
which develop a ``lattice-like'' fifth dimension. This idea
attracted considerable interest.  Our model is quite different, and
very simple: The $SU(N)$ gauge theory  is shown to develop fuzzy
extra dimensions through a standard symmetry breaking mechanism.

\section{The Fuzzy Sphere}
\subsection{Ordinary and Fuzzy spherical harmonics}\label{sec:sphere}

Let us start by recalling how to describe fields on the 2-sphere.
The 2-sphere is a two-dimensional manifold embedded in $\bf{R}^3$,
with a global $SO(3) \sim SU(2)$ isometry group, defined by the
equation
\begin{equation}
x^2_1 + x^2_2 + x^2_3 = R^2
\end{equation}
for a coordinate basis $x_{\hat{a}}$ in $\bf{R}^3$. We define the
coordinates $x_{\hat{a}}$ in terms of the spherical coordinates $y_a
= (\theta, \phi)$ and radius $R$ by,
\begin{eqnarray}
x_1 & = & R \, \textrm{sin} \, \theta \, \textrm{cos} \, \phi, \\
x_2 & = & R \, \textrm{sin} \, \theta \, \textrm{sin} \, \phi, \\
x_3 & = & R \, \textrm{cos} \, \theta,
\end{eqnarray}
which dictates the metric of the 2-sphere,
\begin{equation}
ds^2 = R^2 \, d\theta^2 + R^2 \, \textrm{sin}^2 \theta \, d\phi^2.
\end{equation}
The generators of $SU(2) \sim SO(3)$ are the angular momentum
operators $L_i$,
\begin{equation}
L_{\hat{a}} = -i \varepsilon_{\hat{a}\hat{b}\hat{c}} x_{\hat{b}}
\partial_{\hat{c}}
\end{equation}
In terms of spherical coordinates the angular momentum operators are
\begin{eqnarray}
L_1 & = & \phantom{-} i \, \textrm{sin} \, \phi \,
\frac{\partial}{\partial \theta} + i \, \textrm{cos} \, \phi \,
\textrm{cot} \, \theta \,
\frac{\partial}{\partial \phi}, \\
L_2 & = & -i \, \textrm{cos} \, \phi \, \frac{\partial}{\partial
\theta} + i\, \textrm{sin} \, \phi \, \textrm{cot} \, \theta \,
\frac{\partial}{\partial \phi}, \\
L_3 & = & -i \, \frac{\partial}{\partial \phi},
\end{eqnarray}
which we can summarize as
\begin{equation}\label{eq:killing}
L_{\hat{a}} = -i k^{a}_{\hat{a}} \partial_a
\end{equation}
The metric tensor can also be expressed in terms of the Killing
vectors $k^{a}_{\hat{a}}$ (defined by the above equations) as
\begin{equation}
g^{ab} = \frac{1}{R^2} \, k^{a}_{\hat{a}} k^b_{\hat{a}}.
\end{equation}

We can expand any function on the 2-sphere in terms of the
eigenfunctions of the 2-sphere,
\begin{equation}\label{eq:expand}
a(\theta, \phi) = \sum^{\infty}_{l=0} \sum^l_{m=-l} a_{lm}
Y_{lm}(\theta, \phi),
\end{equation}
where $a_{lm}$ is a complex coefficient and $Y_{lm}(\theta, \phi)$
are the spherical harmonics, which satisfy the equation
\begin{equation}
L^2 Y_{lm} = -R^2 \Delta_{S^2} Y_{lm} = l(l+1) Y_{lm},
\end{equation}
where $\Delta_{S^2}$ is the scalar Laplacian on the 2-sphere
\begin{equation}
\Delta_{S^2} = \frac{1}{\sqrt{g}} \, \partial_a ( g^{ab} \sqrt{g} \,
\partial_b).
\end{equation}
The spherical harmonics have an eigenvalue $\mu \sim l(l+1)$ for
integer \mbox{$l = 0,1, \dots$}, with degeneracy $2l+1$. The
orthogonality condition of the spherical harmonics is
\begin{equation}
\int d\Omega \, Y^{\dag}_{lm} Y^{\phantom{\dag}}_{l'm'} = \delta_{l
l'} \, \delta_{m m'},
\end{equation}
where $d\Omega = \sin \theta \, d\theta d\phi$.

The spherical harmonics can be expressed in terms of the cartesian
coordinates $x_{\hat{a}}$ (with $\hat{a}=1,2,3$) of a unit vector in
$\bf{R}^{3}$,
\begin{equation}\label{eq:spharm}
Y_{lm}(\theta, \phi) = \sum_{\vec{a}} f ^{(lm)} _{\hat{a_{1}} \dots
\hat{a_{l}}} x^{\hat{a_{1}}} \! \dots x^{\hat{a_{l}}}
\end{equation}
where $f ^{(lm)} _{\hat{a_{1}} \dots \hat{a_{l}}}$ is a traceless
symmetric tensor of $SO(3)$ with rank $l$.

Similarly we can expand $N \times N$ matrices on a  sphere as,
\begin{eqnarray}\label{eq:fuzzyexpand}
\hat{a} & = & \sum^{N-1}_{l=0} \sum^l_{m=-l} a_{lm} \hat{Y}_{lm} \\
\hat{Y}_{lm} &
 = & R^{-l}\sum_{\vec{a}} f ^{(lm)} _{\hat{a_{1}} \dots
\hat{a_{l}}} \hat{x}^{\hat{a_{1}}} \! \dots \hat{x}^{\hat{a_{l}}},
\end{eqnarray}
where $\hat{x}_{\hat{a}}=\frac{2R}{\sqrt{N^{2}-1}} \,
X^{(N)}_{\hat{a}}$ are the generators of $SU(2)$ in the
$N$-dimensional representation and $f ^{(lm)}_{\hat{a_{1}} \dots
\hat{a_{l}}}$ is the same tensor as in (\ref{eq:spharm}). The
matrices $\hat{Y}_{lm}$ are known as fuzzy spherical harmonics for
reasons explained in the next subsection. They obey the
orthonormality condition
\begin{equation}
\textrm{Tr}_N \left( \hat{Y}^{\dag}_{lm}
\hat{Y}_{l'm'}^{\phantom{\dag}} \right) = \delta_{l l'} \, \delta_{m
m'}.
\end{equation}
There is an obvious relation between equations (\ref{eq:expand}) and
(\ref{eq:fuzzyexpand}), namely
\begin{equation}\label{eq:map}
\hat{a} = \sum^{N-1}_{l=0} \sum^{l}_{m=-l} a_{lm} \hat{Y}_{lm} \to \
a(\theta, \phi) = \sum^{N-1}_{l = 0} \sum^l_{m = -l} a_{lm}
Y_{lm}(\theta, \phi).
\end{equation}
Notice that the expansion in spherical harmonics is truncated at
$N-1$ reflecting the finite number of degrees of freedom in the
matrix $\hat{a}$. This allows the consistent definition of a matrix
approximation of the sphere known as fuzzy sphere.

\subsection{The Matrix Geometry of the fuzzy sphere} According to
the above discussion the fuzzy sphere \cite{Mad, Madore} is a matrix
approximation of the usual sphere $S^2$. The algebra of functions on
$S^2$ (for example spanned by the spherical harmonics) as explained
in the previous section is truncated at a given frequency and thus
becomes finite dimensional. The truncation has to be consistent with
the associativity of the algebra and this can be nicely achieved
relaxing the commutativity property of the algebra. The fuzzy sphere
is the ``space'' described by this non-commutative algebra. The
algebra itself is that of $N\times N$ matrices. More precisely, the
algebra of functions on the ordinary sphere can be generated by the
coordinates of {\bf{R}}$^3$ modulo the relation $
\sum_{\hat{a}=1}^{3} {x}_{\hat{a}}{x}_{\hat{a}} =r^{2}$. The fuzzy
sphere $S^2_{N}$ at fuzziness level $N-1$ is the non-commutative
manifold whose coordinate functions $i {X}_{\hat{a}}$ are $N \times
N$ hermitian matrices proportional to the generators of the
$N$-dimensional representation of $SU(2)$. They satisfy the
condition $ \sum_{\hat{a}=1}^{3} X_{\hat{a}} X_{\hat{a}} = \alpha
r^{2}$ and the commutation relations
\begin{equation}
[ X_{\hat{a}}, X_{\hat{b}} ] = C_{\hat{a} \hat{b} \hat{c}}
X_{\hat{c}}~,
\end{equation}
where $C_{\hat{a} \hat{b} \hat{c}}= \varepsilon_{\hat{a} \hat{b}
\hat{c}}/r$ while the proportionality factor $\alpha$ goes as $N^2$
for $N$ large. Indeed it can be proven that for $N\rightarrow
\infty$ one obtains the usual commutative sphere.

On the fuzzy sphere there is a natural $SU(2)$ covariant
differential calculus. This calculus is three-dimensional and the
derivations $e_{\hat{a}}$ along $X_{\hat{a}}$ of a function $ f$ are
given by $e_{\hat{a}}({f})=[X_{\hat{a}}, {f}]\,.\label{derivations}$
Accordingly the action of the Lie derivatives on functions is given
by
\begin{equation}\label{LDA}
{\cal L}_{\hat{a}} f = [{X}_{\hat{a}},f ]~;
\end{equation}
these Lie derivatives satisfy the Leibniz rule and the $SU(2)$ Lie
algebra relation
\begin{equation}\label{LDCR}
[ {\cal L}_{\hat{a}}, {\cal L}_{\hat{b}} ] = C_{\hat{a} \hat{b}
\hat{c}} {\cal L}_{\hat{c}}.
\end{equation}
In the $N \rightarrow \infty$ limit the derivations $e_{\hat{a}}$
become $
e_{\hat{a}} = C_{\hat{a} \hat{b} \hat{c}} x^{\hat{b}}
\partial^{\hat{c}}\,
$ and only in this commutative limit the tangent space becomes
two-dimensional. The exterior derivative is given by
\begin{equation}
d f = [X_{\hat{a}},f]\theta^{\hat{a}}
\end{equation}
with $\theta^{\hat{a}}$ the one-forms dual to the vector fields
$e_{\hat{a}}$,
$<e_{\hat{a}},\theta^{\hat{b}}>=\delta_{\hat{a}}^{\hat{b}}$. The
space of one-forms is generated by the $\theta^{\hat{a}}$'s in the
sense that for any one-form $\omega=\sum_i f_i d h_i \:t_i$ we can
always write
$\omega=\sum_{\hat{a}=1}^3{\omega}_{\hat{a}}\theta^{\hat{a}}$ with
given functions $\omega_{\hat{a}}$ depending on the functions $f_i$,
$h_i$ and $t_i$. The action of the Lie derivatives ${\cal
L}_{\hat{a}}$ on the one-forms $\theta^{\hat{b}}$ explicitly reads
\begin{equation}\label{2.16}
{\cal L}_{\hat{a}}(\theta^{\hat{b}}) =  C_{\hat{a}\hat{b}\hat{c}}
\theta^{\hat{c}}~.
\end{equation}
On a general one-form $\omega=\omega_{\hat{a}}\theta^{\hat{a}}$ we
have $ {\cal L}_{\hat{b}}\omega={\cal
L}_{\hat{b}}(\omega_{\hat{a}}\theta^{\hat{a}})=
\left[X_{\hat{b}},\omega_{\hat{a}}\right]\theta^{\hat{a}}-\omega_{\hat{a}}C^{\hat{a}}_{\
\hat{b} \hat{c}}\theta^{\hat{c}} $ and therefore
\begin{equation}
({\cal
L}_{\hat{b}}\omega)_{\hat{a}}=\left[X_{\hat{b}},\omega_{\hat{a}}\right]-
\omega_{\hat{c}}C^{\hat{c}}_{\ \hat{b}  \hat{a}}~;\label{fund}
\end{equation}
this formula will be fundamental for formulating the CSDR principle
on fuzzy cosets.

The differential geometry on  the product space Minkowski times
fuzzy sphere, $M^{4} \times S^2_{N}$, is easily obtained from that
on $M^4$ and on $S^2_N$. For example a one-form $A$ defined on
$M^{4} \times S^2_{N}$ is written as
\begin{equation}\label{oneform}
A= A_{\mu} dx^{\mu} + A_{\hat{a}} \theta^{\hat{a}}
\end{equation}
with $A_{\mu} =A_{\mu}(x^{\mu}, X_{\hat{a}} )$ and $A_{\hat{a}}
=A_{\hat{a}}(x^{\mu}, X_{\hat{a}} )$.

One can also introduce spinors on the fuzzy sphere and study the Lie
derivative on these spinors. Although here we have sketched the
differential geometry on the fuzzy sphere,  one can study other
(higher-dimensional) fuzzy spaces (e.g. fuzzy $CP^M$) and with
similar techniques their differential geometry.

\section{Dimensional Reduction of Fuzzy Extra Dimensions}

\subsection{Actions in higher dimensions seen as four-dimensional
actions (Expansion in Kaluza-Klein modes)} First we consider on
$M^{4} \times (S/R)_{F}$ a non-commutative gauge theory with gauge
group $G=U(P)$ and examine its four-dimensional interpretation.
$(S/R)_{F}$ is a fuzzy coset, for example the fuzzy sphere
$S^{2}_{N}$. The action is
\begin{equation}\label{formula8}
{\cal A}_{YM}={1\over 4g^{2}} \int d^{4}x\, kTr\, tr_{G}\,
F_{MN}F^{MN},
\end{equation}
where $kTr$ denotes integration over the fuzzy coset $(S/R)_F\,$
described by $N\times N$ matrices; here the parameter $k$ is related
to the size of the fuzzy coset space. For example for the fuzzy
sphere we have $r^{2} = \sqrt{N^{2}-1}\pi k$ \cite{Madore}. In the
$N\rightarrow \infty$ limit $kTr$ becomes the usual integral on the
coset space. For finite $N$, $Tr$ is a good integral because it has
the cyclic property $Tr(f_1\ldots f_{p-1}f_p)=Tr(f_pf_1\ldots
f_{p-1})$. It is also invariant under the action of the group $S$,
that is  infinitesimally given by the Lie derivative. In the action
(\ref{formula8}) $tr_G$ is the gauge group $G$ trace. The
higher-dimensional field strength $F_{MN}$, decomposed in
four-dimensional space-time and extra-dimensional components, reads
as follows $(F_{\mu \nu}, F_{\mu \hat{b}}, F_{\hat{a} \hat{b} })\,;$
explicitly the various components of the field strength are given by
\begin{eqnarray}
F_{\mu \nu} &=&
\partial_{\mu}A_{\nu} -
\partial_{\nu}A_{\mu} + [A_{\mu}, A_{\nu}],\\[.3 em]
F_{\mu \hat{a}} &=&
\partial_{\mu}A_{\hat{a}} - [X_{\hat{a}}, A_{\mu}] + [A_{\mu},
A_{\hat{a}}], \nonumber\\[.3 em]
F_{\hat{a} \hat{b}} &=&   [ X_{\hat{a}}, A_{\hat{b}}] - [
X_{\hat{b}}, A_{\hat{a}} ] + [A_{\hat{a}} , A_{\hat{b}} ] -
C^{\hat{c}}_{\ \hat{a} \hat{b}}A_{\hat{c}}.
\end{eqnarray}
Under an infinitesimal $ G $ gauge transformation
$\lambda=\lambda(x^{\mu},X^{\hat{a}})$ we have
\begin{equation}
\delta A_{\hat{a}} = -[ X_{\hat{a}}, \lambda] +
[\lambda,A_{\hat{a}}]~,
\end{equation}
thus $F_{MN}$ is covariant under {local} $G$ gauge transformations:
$F_{MN}\rightarrow F_{MN}+[\lambda, F_{MN}]$. This is an
infinitesimal abelian $U(1)$ gauge transformation if $\lambda$ is
just an antihermitian function of the coordinates $x^\mu,
X^{\hat{a}}$ while it is an infinitesimal non-abelian $U(P)$ gauge
transformation if $\lambda$ is valued in ${\rm{Lie}}(U(P))$, the Lie
algebra of hermitian $P\times P$ matrices. In the following we will
always assume ${\rm{Lie}}(U(P))$ elements to commute with the
coordinates $X^{\hat{a}}$. In fuzzy/non-commutative gauge theory and
in Fuzzy-CSDR a fundamental role is played by the covariant
coordinate,
\begin{equation}
\varphi_{\hat{a}} \equiv X_{\hat{a}} + A_{\hat{a}}~.
\end{equation}
This field transforms indeed covariantly under a gauge
transformation, $
\delta(\varphi_{\hat{a}})=[\lambda,\varphi_{\hat{a}}]~. $ In terms
of $\varphi$ the field strength in the non-commutative directions
reads,
\begin{eqnarray}
F_{\mu \hat{a}} &=&
\partial_{\mu}\varphi_{\hat{a}} + [A_{\mu}, \varphi_{\hat{a}}]=
D_{\mu}\varphi_{\hat{a}},\\[.3 em]
F_{\hat{a} \hat{b}} &=& [\varphi_{\hat{a}}, \varphi_{\hat{b}}] -
C^{\hat{c}}_{\ \hat{a} \hat{b}} \varphi_{\hat{c}}~;
\label{action-CSDR}
\end{eqnarray}
and using these expressions the action reads
\begin{equation}
{\cal A}_{YM}= \int d^{4}x\, Tr\, tr_{G}\,\left( {k\over
4g^{2}}F_{\mu \nu}^{2} + {k\over
2g^{2}}(D_{\mu}\varphi_{\hat{a}})^{2} -
V(\varphi)\right),\label{theYMaction}
\end{equation}
where the potential term $V(\varphi)$ is the $F_{\hat{a} \hat{b}}$
kinetic term (in our conventions $F_{\hat{a} \hat{b}}$ is
antihermitian so that $V(\varphi)$ is hermitian and non-negative)
\begin{eqnarray}\label{pot1}
V(\varphi)&=&-{k\over 4g^{2}} Tr\,tr_G \sum_{\hat{a} \hat{b}}
F_{\hat{a} \hat{b}} F_{\hat{a} \hat{b}}
\nonumber \\
%
&=&-{k\over 4g^{2}} Tr\,tr_G \left( [\varphi_{\hat{a}},
\varphi_{\hat{b}}][\varphi^{\hat{a}}, \varphi^{\hat{b}}] -
4C_{\hat{a} \hat{b} \hat{c}} \varphi^{\hat{a}} \varphi^{\hat{b}}
\varphi^{\hat{c}} + 2r^{-2}\varphi^{2} \right).
\end{eqnarray}
The action (\ref{theYMaction}) is naturally interpreted as an action
in four dimensions. The infinitesimal $G$ gauge transformation with
gauge parameter $\lambda(x^{\mu},X^{\hat{a}})$ can indeed be
interpreted just as an $M^4$ gauge transformation. We write
\begin{equation}
\lambda(x^{\mu},X^{\hat{a}})=\lambda^{\alpha}(x^{\mu},X^{\hat{a}}){\cal
T}^{\alpha} =\lambda^{h, \alpha}(x^{\mu})T^{h}{\cal
T}^{\alpha}~,\label{3.33}
\end{equation}
where ${\cal T}^{\alpha}$ are hermitian generators of $U(P)$,
$\lambda^{\alpha}(x^\mu,X^{\hat{a}})$ are $n\times n$ antihermitian
matrices and thus are expressible as $\lambda(x^\mu)^{\alpha ,
h}T^{h}$, where $T^{h}$ are antihermitian generators of $U(n)$. The
fields $\lambda(x^{\mu})^{\alpha , h}$, with $h=1,\ldots n^2$, are
the Kaluza-Klein modes of $\lambda(x^{\mu}, X^{\hat{a}})^{\alpha}$.
We now consider on equal footing the indices $h$ and $\alpha$ and
interpret the fields on the r.h.s. of (\ref{3.33}) as one field
valued in the tensor product Lie algebra ${\rm{Lie}}(U(n)) \otimes
{\rm{Lie}}(U(P))$. This Lie algebra is indeed ${\rm{Lie}}(U(nP))$
(the $(nP)^2$ generators $T^{h}{\cal T}^{\alpha}$ being $nP\times
nP$ antihermitian matrices that are linear independent). Similarly
we rewrite the gauge field $A_\nu$ as
\begin{equation}
A_\nu(x^{\mu},X^{\hat{a}})=A_{\nu}^{\alpha}(x^{\mu},X^{\hat{a}}){\cal
T}^{\alpha} =A_{\nu}^{h, \alpha}(x^{\mu})T^{h}{\cal T}^{\alpha},
\end{equation}
and interpret it as a ${\rm{Lie}}(U(nP))$ valued gauge field on
$M^4$, and similarly for $\varphi_{\hat{a}}$. Finally $Tr\, tr_{G}$
is the trace over $U(nP)$ matrices in the fundamental
representation.

Up to now we have just performed a ordinary fuzzy dimensional
reduction. Indeed in the commutative case the expression
(\ref{theYMaction}) corresponds to rewriting the initial lagrangian
on $M^4\times S^2$ using spherical harmonics on $S^2$. Here the
space of functions is finite dimensional and therefore the infinite
tower of modes reduces to the finite sum given by $Tr$.
\subsection{Non-trivial Dimensional reduction in the case of Fuzzy
Extra Dimensions} Next we  reduce the number of gauge fields and
scalars in the action (\ref{theYMaction}) by applying the Coset
Space Dimensional Reduction (CSDR) scheme. Since $SU(2)$ acts on the
fuzzy sphere $(SU(2)/U(1))_F$, and more in general  the group $S$
acts on the fuzzy coset $(S/R)_F$, we can state the CSDR principle
in the same way as in the continuum case, i.e. the fields in the
theory must be invariant under the infinitesimal $SU(2)$,
respectively $S$, action up to an infinitesimal gauge transformation
\begin{equation}
{\cal L}_{\hat{b}} \phi = \delta_{W_{\hat{b}}}\phi= W_{\hat{b}}\phi,
\end{equation}
\begin{equation}
{\cal L}_{\hat{b}}A = \delta_{W_{\hat{b}}}A=-DW_{\hat{b}},
\label{csdr}
\end{equation}
where $A$ is the one-form gauge potential $A = A_{\mu}dx^{\mu} +
A_{\hat{a}} \theta^{\hat{a}}$, and $W_{\hat{b}}$ depends only on the
coset coordinates $X^{\hat{a}}$ and (like $A_\mu, A_a$) is
antihermitian. We thus write $W_{\hat{b}}=W_{\hat{b}}^{\alpha}{\cal
T}^{\alpha}, \,\alpha=1,2\ldots P^2,$ where ${\cal  T}^i$ are
hermitian generators of $U(P)$ and $(W_b^i)^\dagger=-W_b^i$, here
${}^\dagger$ is hermitian conjugation on the $X^{\hat{a}}$'s.

In terms of the covariant coordinate $\varphi_{\hat{d}} =X_{\hat{d}}
+ A_{\hat{d}}$ and of
\begin{equation}
\omega_{\hat{a}} \equiv X_{\hat{a}} - W_{\hat{a}}~,
\end{equation}
the CSDR constraints assume a particularly simple form, namely
\begin{equation}\label{3.19}
[\omega_{\hat{b}}, A_{\mu}] =0,
\end{equation}
\begin{equation}\label{eq7}
C_{\hat{b} \hat{d} \hat{e}} \varphi^{\hat{e}} = [\omega_{\hat{b}},
\varphi_{\hat{d}} ].
\end{equation}
In addition we  have a consistency condition  following from the
relation $[{\cal{L}}_{\hat{a}},{\cal{L}}_{\hat{b}}]=
C_{\hat{a}\hat{b}}^{~~\hat{c}}{\cal{L}}_{\hat{c}}$:
\begin{equation}\label{3.17}
[ \omega_{\hat{a}} , \omega_{\hat{b}}] = C_{\hat{a} \hat{b}}^{\ \
\hat{c}} \omega_{c},
\end{equation}
where $\omega_{\hat{a}}$ transforms as $ \omega_{\hat{a}}\rightarrow
\omega'_{\hat{a}} = g\omega_{\hat{a}}g^{-1}. $ One proceeds in a
similar way for the spinor fields \cite{Aschieri:2003vy,
Aschieri:2004vh}.
\subsubsection{Solving the CSDR constraints for
the fuzzy sphere}We consider $(S/R)_{F}=S^{2}_{N}$, i.e. the fuzzy
sphere, and to be definite at fuzziness level $N-1$ ($N \times N$
matrices). We study here the basic example where the gauge group is
$G=U(1)$. In this case the
$\omega_{\hat{a}}=\omega_{\hat{a}}(X^{\hat{b}})$ appearing in the
consistency condition (\ref{3.17}) are $N \times N$ antihermitian
matrices and therefore can be interpreted as elements of
${\rm{Lie}}(U(N))$. On the other hand the $\omega_{\hat{a}}$ satisfy
the commutation relations (\ref{3.17}) of ${\rm{Lie}}(SU(2))$.
Therefore in order to satisfy the consistency condition (\ref{3.17})
we have to embed ${\rm{Lie}}(SU(2))$ in ${\rm{Lie}}(U(N))$. Let
$T^h$ with $h = 1, \ldots ,(N)^{2}$ be the generators of
${\rm{Lie}}(U(N))$ in the fundamental representation, we can always
use the convention $h= (\hat{a} , u)$ with $\hat{a} = 1,2,3$ and $u=
4,5,\ldots, N^{2}$ where the $T^{\hat{a}}$ satisfy the $SU(2)$ Lie
algebra,
\begin{equation}
[T^{\hat{a}}, T^{\hat{b}}] = C^{\hat{a} \hat{b}}_{\ \
\hat{c}}T^{\hat{c}}~.
\end{equation}
Then we define an embedding by identifying
\begin{equation}
 \omega_{\hat{a}}= T_{\hat{a}}.
\label{embedding}
\end{equation}
The constraint (\ref{3.19}), $[\omega_{\hat{b}} , A_{\mu}] = 0$,
then implies that the four-dimensional gauge group $K$ is the
centralizer of the image of $SU(2)$ in $U(N)$, i.e. $$
K=C_{U(N)}(SU((2))) = SU(N-2) \times U(1)\times U(1)~, $$  where the
last $U(1)$ is the $U(1)$ of $U(N)\simeq SU(N)\times U(1)$. The
functions $A_{\mu}(x,X)$ are arbitrary functions of $x$ but the $X$
dependence is such that $A_{\mu}(x,X)$ is ${\rm{Lie}}(K)$ valued
instead of ${\rm{Lie}}(U(N))$, i.e. eventually we have a
four-dimensional gauge potential $A_\mu(x)$ with values in
${\rm{Lie}}(K)$. Concerning the constraint (\ref{eq7}), it is
satisfied by choosing
\begin{equation}
\label{soleasy} \varphi_{\hat{a}}=r \varphi(x) \omega_{\hat{a}}~,
\end{equation}
i.e. the unconstrained degrees of freedom correspond to the scalar
field $\varphi(x)$ which is a singlet under the four-dimensional
gauge group $K$.

The choice (\ref{embedding}) defines one of the possible embedding
of ${\rm{Lie}}(SU(2))$ in ${\rm{Lie}}(U(N))$. For example we could
also embed ${\rm{Lie}}(SU(2))$ in ${\rm{Lie}}(U(N))$ using the
irreducible $N$-dimensional rep. of $SU(2)$, i.e. we could identify
$\omega_{\hat{a}}= X_{\hat{a}}$. The constraint (\ref{3.19}) in this
case implies that the four-dimensional gauge group is $U(1)$ so that
$A_\mu(x)$ is $U(1)$ valued. The constraint (\ref{eq7}) leads again
to the scalar singlet $\varphi(x)$.

In general, we start with a $U(1)$ gauge theory on $M^4\times
S^2_N$. We solve the CSDR constraint (\ref{3.17}) by embedding
$SU(2)$ in $U(N)$. There exist $p_{N}$ embeddings, where $p_N$ is
the number of ways one can partition the integer $N$ into a set of
non-increasing positive integers \cite{Mad}. Then the constraint
(\ref{3.19}) gives the surviving four-dimensional gauge group. The
constraint (\ref{eq7}) gives the surviving four-dimensional scalars
and eq. (\ref{soleasy}) is always a solution but in general not the
only one. By setting $\phi_{\hat{a}}=\omega_{\hat{a}}$ we obtain
always a minimum of the potential. This minimum is given by the
chosen embedding of $SU(2)$ in $U(N)$.

An important point that we would like to stress here is the question
of the renormalizability of the gauge theory defined on $M_4 \times
(S/R)_F$. First we notice that the theory exhibits certain features
so similar to a higher-dimensional gauge theory defined on $M_4
\times S/R$ that naturally it could be considered as a
higher-dimensional theory too. For instance the isometries of the
spaces $M_4 \times S/R$ and $M_4 \times (S/R)_F$ are the same. It
does not matter if the compact space is fuzzy or not. For example in
the case of the fuzzy sphere, i.e. $M_4 \times S^2_N$, the
isometries are $SO(3,1) \times SO(3)$ as in the case of the
continuous space, $M_4 \times S^2$. Similarly the coupling of a
gauge theory defined on $M_4 \times S/R$ and on $M_4 \times (S/R)_F$
are both dimensionful and have exactly the same dimensionality. On
the other hand the first theory is clearly non-renormalizable, while
the latter is renormalizable (in the sense that divergencies can be
removed by a finite number of counterterms). So from this point of
view one finds a partial justification of the old hopes for
considering quantum field theories on non-commutative structures. If
this observation can lead  to finite theories too, it remains as an
open question.

\section{Dynamical Generation of Extra Dimensions }

Let us now discuss a further development \cite{Aschieri:2006uw} of
these ideas,
 which addresses in detail the questions of
 quantization and renormalization. This leads to a slightly
modified model with an extra term in the potential, which
dynamically selects a unique (nontrivial) vacuum out of the many
possible CSDR solutions, and moreover generates a magnetic flux on
the fuzzy sphere. It also allows to show that the full tower of
Kaluza-Klein modes is generated on $S^2_N$.

\subsection{ The four dimensional action }

We start with a $SU(N)$ gauge theory on four dimensional Minkowski
space $M^4$ with coordinates $y^\mu$, $\mu = 0,1,2,3$.  The action
under consideration is
\begin{equation}
{\cal S}_{YM}= \int d^{4}y\, Tr\,\left( \frac{1}{4g^{2}}\, F_{\mu
\nu}^\dagger F_{\mu \nu} + (D_{\mu}\phi_{{a}})^\dagger
D_{\mu}\phi_{{a}}\right) - V(\phi) \label{the4daction}
\end{equation}
where $A_\mu$ are $su(N)$-valued gauge fields, $D_\mu =
\partial_\mu + [A_\mu,.]$, and
$$ \phi_{{a}} = - \phi_{{a}}^\dagger ~, \qquad a=1,2,3 $$ are 3
antihermitian scalars in the adjoint of $SU(N)$, $$ \phi_{{a}} \to
U^\dagger \phi_{{a}} U $$ where $U = U(y) \in SU(N)$. Furthermore,
the $\phi_a$ transform as vectors of an additional global $SO(3)$
symmetry. The potential $V(\phi)$ is taken to be the most general
renormalizable action invariant under the above symmetries, which is
\begin{eqnarray}
V(\phi) &=& Tr\, \left( g_1 \phi_a\phi_a \phi_b\phi_b +
g_2\phi_a\phi_b\phi_a \phi_b - g_3 \varepsilon_{a b c} \phi_a \phi_b
\phi_c + g_4\phi_a \phi_a \right) \nonumber\\ && + \frac{g_5}{N}\,
Tr(\phi_a \phi_a)Tr(\phi_b \phi_b) + \frac{g_6}{N} Tr(\phi_a
\phi_b)Tr(\phi_a \phi_b) +g_7. \label{pot}
\end{eqnarray}
This may not look very transparent at first sight, however it can be
written in a very intuitive way. First, we make the scalars
dimensionless by rescaling $$ \phi'_a = R\; \phi_a, $$ where $R$ has
dimension of length; we will usually suppress $R$ since it can
immediately be reinserted, and drop the prime from now on.  Now
observe that for a suitable choice of $R$, $$
R = \frac{2 g_2}{g_3}, 
\label{Radius} $$ the potential can be rewritten as
\begin{equation}
V(\phi)=Tr ( a^2 \phi_a\phi_a + \tilde b\, )^2 + c +\frac {1}{\tilde
g^2}\, F_{ab}^\dagger F_{ab}\ + \frac{h}{N}\, g_{ab} g_{ab}
\label{V-general-2}
\end{equation}
for suitable constants $a,b,c,\tilde g,h$, where
\begin{eqnarray}
F_{{a}{b}} &=& [\phi_{{a}}, \phi_{{b}}] -
\varepsilon_{abc} \phi_{{c}}\, = \varepsilon_{abc} F_c , \nonumber\\
\tilde b &=& b + \frac{d}{N} \, Tr(\phi_a \phi_a), \nonumber\\
g_{ab} &=& Tr(\phi_a \phi_b). \label{const-def}
\end{eqnarray}
We will omit $c$ from now.  The potential is clearly positive
definite provided $$ a^2 = g_1+g_2 >0, \qquad \frac 2{\tilde g^2} =
- g_2 >0, \qquad h \geq 0, $$ which we assume from now on.  Here
$\tilde b = \tilde b(y)$ is a scalar, $g_{ab} = g_{ab}(y)$ is a
symmetric tensor under the global $SO(3)$, and $F_{ab}=F_{ab}(y)$ is
a $su(N)$-valued antisymmetric tensor field which will be
interpreted as field strength in some dynamically generated extra
dimensions below.  In this form, $V(\phi)$ looks like the action of
Yang-Mills gauge theory on a fuzzy sphere in the matrix formulation
\cite{Steinacker:2003sd,Steinacker:2004yu,Carow-Watamura:1998jn,
Presnajder:2003ak}. It differs from the potential in
(\ref{action-CSDR}) only by the presence of the first term $a^2
(\phi_a\phi_a + \tilde b)^2$, which is strongly suggested
 by renormalization.
In fact it is necessary for the interpretation as pure YM action,
and we will see that it is very welcome on physical grounds since it
dynamically determines and stabilizes a vacuum, which can be
interpreted as extra-dimensional fuzzy sphere. In particular, it
removes unwanted flat directions.

\subsection{Emergence of extra dimensions and the fuzzy sphere}
\label{sec:emergence}

The vacuum of the above model is given by the minimum of the
potential (\ref{pot}). It turns out \cite{Aschieri:2006uw} that
there are essentially only 2 types of vacua:
\begin{enumerate}
\item {\em Type I vacuum}
$\,\,\phi_a = \alpha\, X_a^{(N)} \otimes 1_{n}$ with low-energy
gauge group $SU(n)$, and
\item {\em Type I vacuum}
$\,\,\phi_a = \left(\begin{array}{cc}\alpha_1\, X_a^{(N_1)}\otimes
1_{n_1} & 0 \\ 0 & \alpha_2\,X_a^{(N_2)}\otimes 1_{n_2}
             \end{array}\right)$,
with low-energy gauge group $SU(n_1)\times SU(n_2) \times U(1)$.
\end{enumerate}
Again, the $X_a^{(N)}$ are interpreted as coordinate functions of a
fuzzy sphere $S^2_{N}$, and the ``scalar'' action
$$ S_{\phi} = Tr V(\phi) = Tr\Big(a^2 (\phi_a\phi_a + \tilde b)^2 +
\frac 1{\tilde g^2}\, F_{ab}^\dagger F_{ab}\Big) \label{S-YM2}$$ for
$N \times N$ matrices $\phi_a$ is precisely the action for a $U(n)$
Yang-Mills theory on $S^2_{N}$ with coupling $\tilde g$, as shown in
\cite{Steinacker:2003sd}. In fact, the new term $(\phi_a\phi_a +
\tilde b)^2$ is essential for this interpretation, since it
stabilizes the vacuum $\phi_a = X_a^{(N)}$ and gives a large mass to
the extra ``radial'' scalar field which otherwise arises.  The
fluctuations of $\phi_a = X_a^{(N)} + A_a$ then provide the
components $A_a$ of a higher-dimensional gauge field $A_M = (A_\mu,
A_a)$, and the action can be interpreted as YM theory on the
6-dimensional space $M^4 \times S^2_{N}$, with gauge group depending
on the particular vacuum.  We therefore interpret the vacuum as
describing dynamically generated extra dimensions in the form of a
fuzzy sphere $S^2_{N}$. This geometrical interpretation can be fully
justified
 by working out the spectrum of Kaluza-Klein
modes.  The effective low-energy theory is then given by the zero
modes on $S^2_{N}$. This approach provides a clear dynamical
selection of the geometry due to the term $(\phi_a\phi_a + \tilde
b)^2$ in the action.

Perhaps the most remarkable aspect of this model is that the
geometric interpretation and the corresponding low-energy degrees of
freedom depend in a nontrivial way on the parameters of the model,
which are running under the RG group. Therefore the massless degrees
of freedom and their geometrical interpretation depend on the energy
scale. In particular, the low-energy gauge group generically turns
out to be $SU(n_1) \times SU(n_2)\times U(1)$ or $SU(n)$, while
 gauge groups which are
products of more than two simple components (apart from $U(1)$) do
not seem to occur. The values of $n_1$ and $n_2$ are determined
dynamically, and may well be small such as 3 and 2.

It is interesting to examine  the running of the coupling constants
under the RG. $R$ turns out to run only logarithmically, implies
that the scale of the internal spheres is only mildly affected by
the RG flow. However, $\tilde b$ is running essentially
quadratically, hence is generically large. This is quite welcome
here: starting with some large $N$, $\tilde{b} \approx
C_2(\tilde{N})$ must indeed be large in order to lead to the
geometric interpretation discussed above. Hence the problems of
naturalness or fine-tuning appear to be rather mild here.

A somewhat similar model has been studied  in
\cite{Andrews:2005cv,Andrews:2006aw}, which realizes deconstruction
and a ``twisted'' compactification of an extra fuzzy sphere based on
a supersymmetric gauge theory. Our model is different and does not
require supersymmetry, leading to a much richer pattern of symmetry
breaking and effective geometry. For other relevant work see e.g.
\cite{Madore:1992ej}.

The dynamical formation of fuzzy spaces found here is also related
to recent work studying the emergence of stable submanifolds in
modified IIB matrix models. In particular, previous studies based on
actions for fuzzy gauge theory different from ours generically only
gave results corresponding to $U(1)$ or $U(\infty)$ gauge groups,
see e.g. \cite{Azuma:2004ie,Azuma:2005bj,Azuma:2004zq} and
references therein. The dynamical generation of a nontrivial index
on noncommutative spaces has also been observed in
\cite{Aoki:2004sd,Aoki:2006zi} for different models.

Our mechanism may also be very interesting in the context of the
recent observation \cite{Abel:2005rh} that extra dimensions are very
desirable for the application of noncommutative field theory to
particle physics. Other related recent work discussing the
implications of the higher-dimensional point of view on symmetry
breaking and Higgs masses can be found in
\cite{Lim:2006bx,Dvali:2001qr,Antoniadis:2002ns,Scrucca:2003ra}.
These issues could now be discussed within a renormalizable
framework.

\section{Discussion and Conclusions}
Non-commutative Geometry has been regarded as a promising framework
for obtaining finite quantum field theories and  for regularizing
quantum field theories. In general quantization of field theories on
non-commutative spaces has turned out to be much more difficult and
with less attractive ultraviolet features than expected
see however ref. \cite{Grosse:2004ik}, and ref. \cite{Steinacker}.
Recall also that non-commutativity is not the only suggested tool
for constructing finite field theories. Indeed four-dimensional
finite gauge theories have been constructed in ordinary space-time
and not only those which are ${\cal N} = 4$ and ${\cal N} = 2$
supersymmetric, and most probably phenomenologically uninteresting,
but also chiral ${\cal N} = 1$ gauge theories \cite{Kapetanakis:vx}
which already have been successful in predicting the top quark mass
and have rich phenomenology that could be tested in future colliders
\cite{Kapetanakis:vx,Kubo:1994bj}. In the present work we have not
adressed the finiteness of non-commutative quantum field theories,
rather we have used non-commutativity to produce, via Fuzzy-CSDR,
new particle models from  particle models on $M^4\times (S/R)_F$.

A major difference between fuzzy and ordinary SCDR is that in the
fuzzy case one always embeds $S$ in the gauge group $G$ instead of
embedding just $R$ in $G$. This is due to the fact that the
differential calculus on the fuzzy coset space is based on $dim S$
derivations instead of the restricted $dim S - dim R$ used in the
ordinary one.  As a result the four-dimensional gauge group $H =
C_G(R)$ appearing in the ordinary CSDR after the geometrical
breaking and before the spontaneous symmetry breaking due to the
four-dimensional Higgs fields does not appear in the Fuzzy-CSDR. In
Fuzzy-CSDR the spontaneous symmetry breaking mechanism takes already
place by solving the Fuzzy-CSDR constraints. The four-dimensional
potential has the typical ``maxican hat'' shape, but it appears
already spontaneously broken. Therefore in four dimensions appears
only the physical Higgs field that survives after a spontaneous
symmetry breaking. Correspondingly in the Yukawa sector of the
theory we have the results of the spontaneous symmetry breaking,
i.e. massive fermions and Yukawa interactions among fermions and the
physical Higgs field. Having massive fermions in the final theory is
a generic feature of CSDR when $S$ is embedded in $G$
\cite{Kapetanakis:hf}. We see that if one would like to describe the
spontaneous symmetry breaking of the SM in the present framework,
then one would be naturally led to large extra dimensions.

A fundamental difference between the ordinary CSDR and its fuzzy
version is the fact that a non-abelian gauge group $G$ is not really
required in high dimensions. Indeed  the presence of a $U(1)$ in the
higher-dimensional theory is enough to obtain non-abelian gauge
theories in four dimensions.


In a further development, we have presented a renormalizable four
dimensional $SU(N)$ gauge theory with a suitable multiplet of
scalars, which dynamically develops fuzzy extra dimensions that form
a fuzzy sphere. The model can then be interpreted as 6-dimensional
gauge theory, with gauge group and geometry depending on the
parameters in the original Lagrangian. We explicitly find the tower
of massive Kaluza-Klein modes, consistent with an interpretation as
compactified higher-dimensional gauge theory, and determine the
effective compactified gauge theory. This model has a unique vacuum,
with associated geometry and low-energy gauge group depending only
on the parameters of the potential.

There are many remarkable aspects of this model.  First, it provides
an extremely simple and geometrical mechanism of dynamically
generating extra dimensions, without relying on subtle dynamics such
as fermion condensation and particular Moose- or Quiver-type arrays
of gauge groups and couplings, such as in \cite{Arkani-Hamed:2001ca}
and following work. Rather, our model is based on a basic lesson
from noncommutative gauge theory, namely that noncommutative or
fuzzy spaces can be obtained as solutions of matrix models. The
mechanism is quite generic, and does not require fine-tuning or
supersymmetry. This provides in particular a realization of the
basic ideas of compactification and dimensional reduction within the
framework of renormalizable quantum field theory. Moreover, we are
essentially considering a large $N$ gauge theory, which should allow
to apply the analytical techniques developed in this context.

In particular, it turns out that the generic low-energy gauge group
is given by $SU(n_1) \times SU(n_2)\times U(1)$ or $SU(n)$, while
 gauge groups which are
products of more than two simple components (apart from $U(1)$) do
not seem to occur in this model. The values of $n_1$ and $n_2$ are
determined dynamically. Moreover,  a magnetic flux is induced in the
vacua with non-simple gauge group, which
 is very interesting in the
context of fermions, since internal fluxes naturally lead to chiral
massless fermions. This will be studied in detail elsewhere.

There is also an intriguing analogy between our toy model and string
theory, in the sense that as long as $a=0$, there are a large number
of possible vacua (given by all possible partitions)
corresponding to compactifications, with no dynamical selection
mechanism to choose one from the other. Remarkably this analog of
the ``string vacuum problem'' is simply solved by adding a term to
the action.

\section*{Acknowledgements}
We would like to thank C. Bachas, T. Grammatikopoulos, H. Grosse and
B. Jurco for discussions. GZ would like to thank the organizers for
the warm hospitality. This work is supported by the EPEAEK
programmes \foreignlanguage{greek}{Πυθαγόρας} and co-founded by the
European Union (75 \%) and the Hellenic State (25 \%) and in part by
the European Commission under the Research and Training Network
contract MRTN-CT-2004-503369; PM is supported by the Hellenic State
Scholarship Foundation (I.K.Y) and by the program
\foreignlanguage{greek}{Πυθαγόρας} (89194).

\vspace{2cm} 


\end{document}